# Unexplored reactivity of $(S_n)^{2-}$ Oligomers with transition metals in low-temperature solid-state reactions

*Shunsuke Sasaki, Melanie Lesault, Elodie Grange, Etienne Janod, Benoît Corraze, Sylvian Cadars, Maria Teresa Caldes, Catherine Guillot-Deudon, Stéphane Jobic\* and Laurent Cario\**

*Institut des Matériaux Jean Rouxel (IMN), Université de Nantes, CNRS, 2 rue de la Houssinière, BP 32229, 44322 Nantes Cedex 3, France.*

\*E-mail. Stephane.Jobic@cnrs-imn.fr, Laurent.Cario@cnrs-imn.fr

## ABSTRACT

Chalcogenides (Q = S, Se, Te), one of the most important family of materials in solid-state chemistry, differ from oxides by their ability to form covalently-bonded $(Q_n)^{2-}$ oligomers. Each chalcogen atom within such entity fulfills the octet rule by sharing electrons with other chalcogen atoms but some antibonding levels are vacant. This makes these oligomers particularly suited for redox reactions in solid state, namely towards elemental metals with a low redox potential that may be oxidized. We recently used this strategy to design, at low temperature and in an orientated manner, materials with 2D infinite layers through the topochemical insertion of copper into preformed precursors containing $(S_2)^{2-}$ and/or $(Se_2)^{2-}$ dimers (*i.e.* $La_2O_2S_2$, $Ba_2F_2S_2$ and $LaSe_2$). Herein we extend the validity of the concept to the redox activity of $(S_2)^{2-}$ and $(S_3)^{2-}$ oligomers towards 3d transition metal elements (Cu, Ni, Fe) and highlight the strong relationship between the structures of the precursors, $BaS_2$ and $BaS_3$, and the products, $BaCu_2S_2$, $BaCu_4S_3$, $BaNiS_2$ and $BaFe_2S_3$. Clearly, beyond the natural interest for the chemical reactivity of oligomers to generate compounds, this soft chemistry route may conduct to the rational conception of materials with a predicted crystal structure.

Chalcogenides has been established as one of the most important field in solid-state chemistry, owing to the richness of their chemical compositions, the complexity of their structural arrangements and their numerous applications (*e.g.* solar cells,[1] thermoelectricity,[2] non-volatile memories[3]). Sulfides, selenides, and tellurides received also much attention for their electronic instabilities,[4] and were recently spotlighted for their optical and electronic properties when prepared as two-dimensional nanosheets.[5-6] From a chemical point of view, the uniqueness of chalcogenides compared to oxides arises from their ability to stabilize a large series of discrete $Q_n$ oligomers or infinite polymerized $Q_n$ network.[7-9] The reactive flux method led, for example, to the stabilization of a myriad of new materials with a large variety of $Q_n$ oligomers (*n* up to *ca.* 10) that may coexist in the same framework.[10] The occurrence of these Q-Q bonds affect drastically the electronic structure and therefore the physical properties[11-13] of the materials in which they are embedded.[14-16] But the presence of Q-Q bonds also confers on these materials a remarkable chemical reactivity in the solid state.[17-20] In that respect, we recently demonstrated that the redox activity of $(Q_2)^{2-}$ pairs (Q = S, Se) could be valuable to design and synthesize layered transition metal chalcogenides in a topochemical manner.[21] We found that the low-temperature solid-solid reactions between elemental copper and precursors containing $(Q_2)^{2-}$ pairs sandwiched between redox-inactive cationic layers trigger the formation of layered quaternary and ternary compounds thanks to the insertion of copper and the Q-Q bond cleavage.

Notably, this novel topochemical route can be written in a general way as follows:

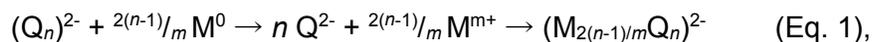

$(Q_n)^{2-} + {}^{2(n-1)}/_m M^0 \rightarrow n\ Q^{2-} + {}^{2(n-1)}/_m M^{m+} \rightarrow (M_{2(n-1)/m}Q_n)^{2-}$   (Eq. 1),

where M = transition metal and Q = S, Se, Te.
Accordingly, it should work with various transition metals and low-dimensional polychalcogenides species as long as the electron transfer from the metal to the oligomer is feasible (see Fig. 1a). It holds therefore the promise to become a versatile synthetic methodology to explore novel transition metal chalcogenides with unconventional properties and a low dimensional structure inherited from the precursor (see Fig. 1b). To evidence such a general applicability, we showcase the reactivity of $BaS_2$ and $BaS_3$ with $(S_2)^{2-}$ dimer and $(S_3)^{2-}$ trimers towards the insertion of copper, nickel or iron. In particular, we demonstrate the ability of the method to stabilize, in soft conditions, $BaNiS_2$, a promising material for spintronic due to its Rashba spin-orbit coupling,[22] and $BaFe_2S_3$, the first spin-ladder iron-based superconductor.[23]

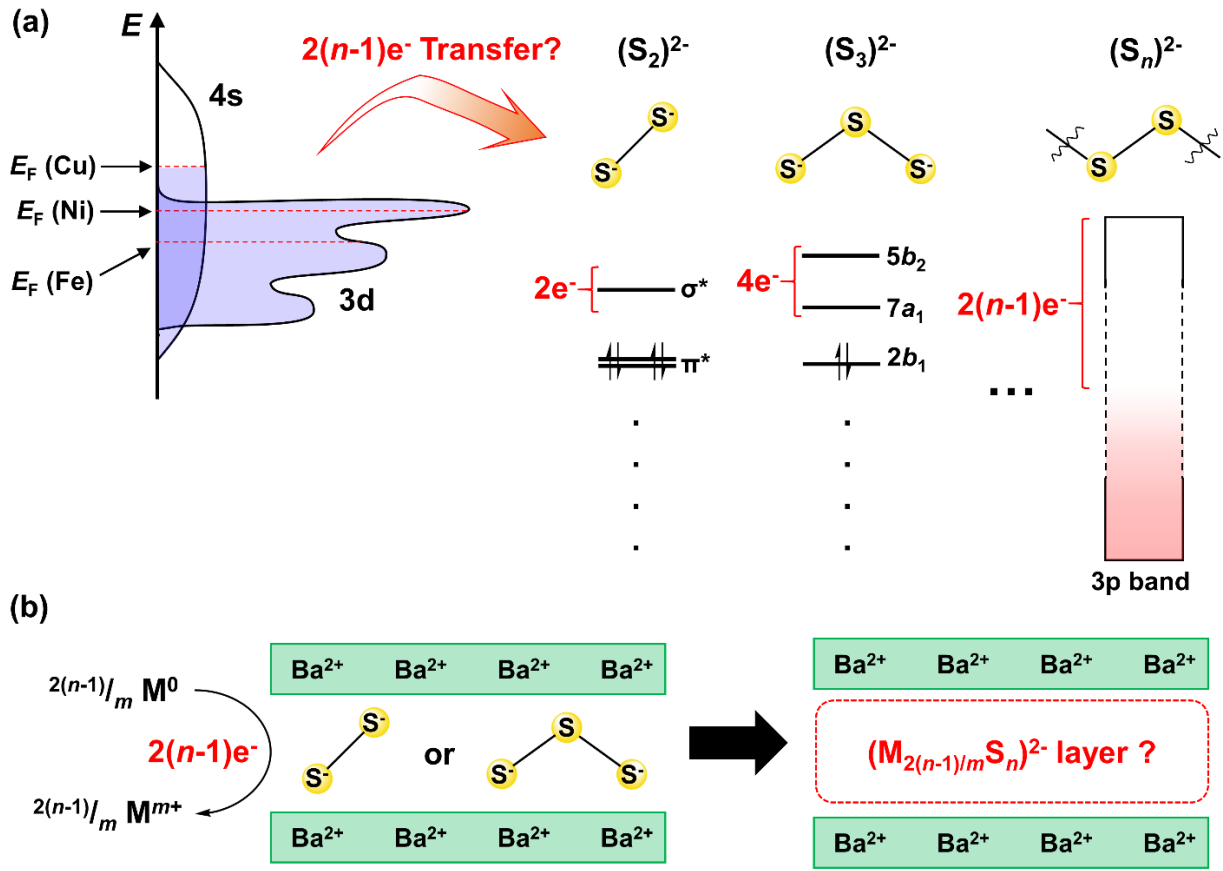

**Fig. 1**. (a) Schematic energy diagram of the redox chemistry between transition metals and sulfur oligomers. The simplistic density of states depicting band filling of 3d metals is adapted from reference 24. (b) Insertion of transition metals triggered by the redox reaction with polychalcogenides leading creation of a two-dimensional transition metal chalcogenide layer sandwiched between "inert" cationic layers.

Barium sulfides are known to exhibit several polysulfide species $BaS_n$ ($n$ = 2 - 5) although tetramers and pentamers are very difficult to isolate.[25] Herein $BaS_2$ and $BaS_3$ have been chosen to examine the reactivity of $(S_2)^{2-}$ and $(S_3)^{2-}$ oligomers towards Cu, Ni and Fe. $BaS_2$ (SG: *C2/c*) and $BaS_3$ (SG: *P-42$_1$m*) materials can be regarded as the 2D assembly of $(S_2)^{2-}$ or $(S_3)^{2-}$ discrete entities that alternate along the stacking axis with $Ba^{2+}$ cationic layers (Fig. 2a, b). Intramolecular S-S distances are 2.12 Å for $(S_2)^{2-}$ dimers and 2.08 Å for $(S_3)^{2-}$ trimers,[26] corresponding to typical values for S-S single bonds.[27] As calculated in our previous report,[28] such $(S_2)^{2-}$ dimers feature molecular orbitals common to homoatomic dimers of p-block elements (Fig. 1a). In the same way our DFT calculation of $BaS_3$ demonstrates that its density of states are characterized clearly by molecular orbitals of $(S_3)^{2-}$ trimers, which accord well with the electronic structure of the isolated

S$_3$ molecule (Fig. S1).[29] Accordingly, BaS$_2$ and BaS$_3$ should be able to accommodate up to 2 and 4 donated electrons per formula unit to yield the closed shell S$^{2-}$ species.

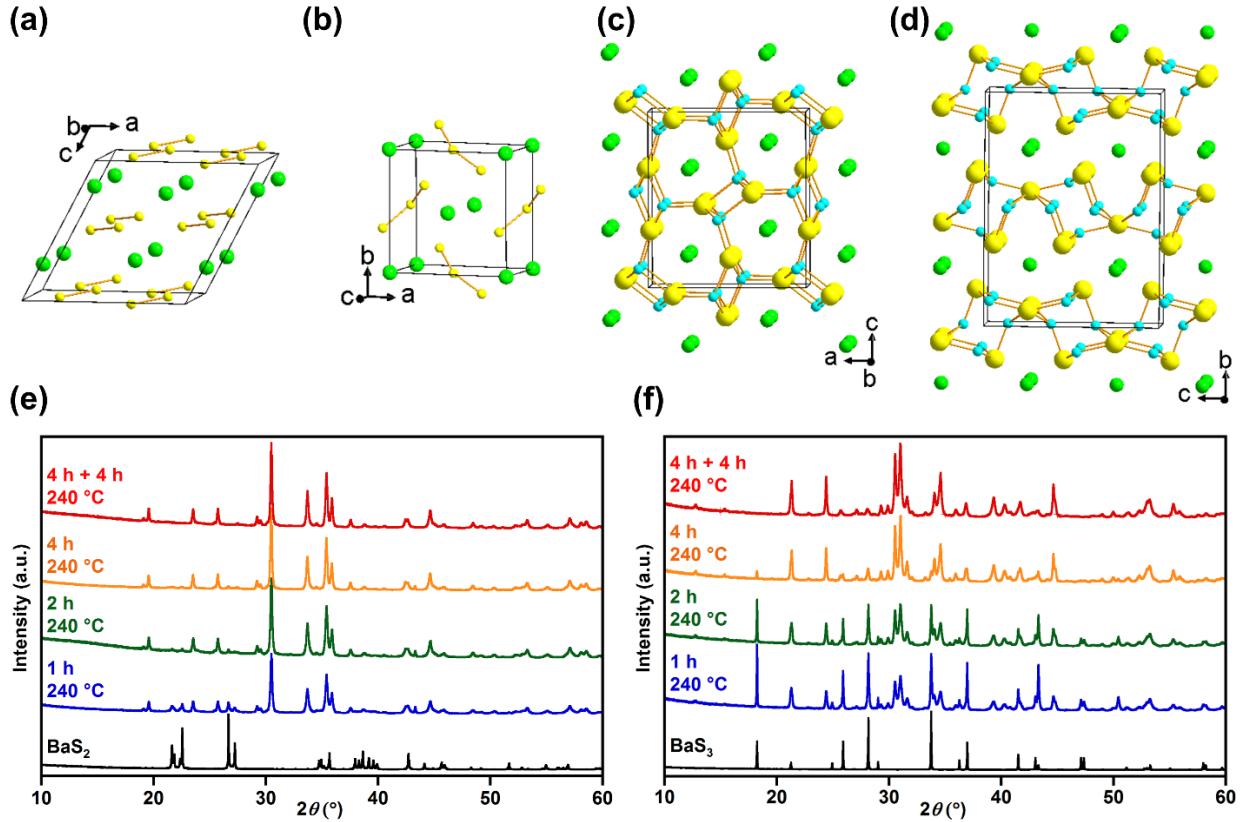

**Fig. 2**. Structures of (a) BaS$_2$, (b) BaS$_3$, (c) α-BaCu$_2$S$_2$ and (d) β-BaCu$_4$S$_3$. Barium, copper and sulfur atoms are represented by lime, cyan and yellow spheres. (e), (f) XRD patterns of the BaS$_2$ + 2Cu and BaS$_3$ + 4Cu mixtures after the thermal treatments at 240 °C for 1, 2, 4 and 4 + 4 h, respectively, respectively.

Let us first consider the incorporation of Cu into BaS$_2$ and BaS$_3$. Following Eq. 1, the BaS$_2$:Cu = 1:2 (mol/mol) mixture was heated at low temperature in a sealed pyrex tube to initiate the insertion of copper in the host lattice. Indeed, X-ray diffraction (XRD) evidenced that peaks of the BaS$_2$ precursor disappeared within few hours at 240 °C while new peaks concomitantly emerged (Fig. 2e, See Fig. S2-S3 for the detail) that were all assigned to α-BaCu$_2$S$_2$ (SG: Pnma) (Fig. 2c),[30] i.e. the low-temperature form known to convert into the high-temperature form (β-BaCu$_2$S$_2$, SG: I4/mmm) above 540 °C.[31] In particular, after annealing at 240 °C for 4h subsequent to a manual milling step, pure BaCu$_2$S$_2$ was obtained. The same synthesis route was applied to the BaS$_3$:Cu = 1:4 (mol/mol) mixture. The heating at low temperature quickly triggers the conversion of BaS$_3$ into BaCu$_4$S$_3$ (Fig. 2d), again without detection of intermediates or impurity phases (Fig. 2f, See Fig. S4 for the detail). According to our Rietveld analysis (See Table S2, Fig. S5 and S6), conversion of $BaS_3$ into $BaCu_4S_3$ was slower than $BaS_2$ into $BaCu_2S_2$. Moreover, unlike the synthesis of $BaCu_2S_2$, $BaCu_4S_3$ was obtained as two polymorphs; the molar ratio between the low-temperature phase (α phase) and the high-temperature one (β phase) being equal to *ca.* 2 : 3 regardless the duration of the heating treatment (Fig. S7). The two polymorphs turn out to be structurally very similar, the β- (SG: Cmcm, Fig. 2d) and α- (SG: Pnma) phase exhibiting flat and corrugated $Ba^{2+}$ and $Cu^+$ layers, respectively.[32] Since its reaction temperature ($T$ = 240 °C) was far lower than the α → β phase transition temperature of 640 ± 10 °C,[32] β phase was formed as a metastable compound. This fact supports the practicability of our soft chemistry route as a novel tool to explore metastable phases. Also Raman peaks ascribed to the normal modes of $(S_2)^{2-}$ dimers[33-34] in $BaS_2$ and $(S_3)^{2-}$ trimers[35] in $BaS_3$ have disappeared in the course of the reaction with copper (Fig. S8) while photoluminescence of the products masked small Raman signals. In any case, these synthesis results with $BaS_2$ and $BaS_3$ as precursors clearly demonstrate the redox activity of not only $(S_2)^{2-}$ dimers but also $(S_3)^{2-}$ trimers towards the insertion of copper. This type of intercalation reactions is likely to occur with other oligomers and might also explain the stabilization of $KCu_4S_3$ via heating of a $K_2S_5$ + Cu mixture while $K_2S_5$ used as a flux was not yet melted.[36]

So far, the reactivity of metal towards holes at empty antibonding levels of $(S_n)^{2-}$ polysulfides has been spotlighted for copper only.[21] Herein we examine the reactivity of $(S_n)^{2-}$ oligomers toward open-shell 3d transition metals, namely Ni and Fe, whose work functions are comparable with copper (Ni (5.2 eV) > Cu (4.64 eV) > Fe (4.5 eV)[37] but stable as divalent cations in chalcogenides.

Practically, Ni was initially mixed to $BaS_2$ in 1:1 ratio, and then heated at 340°C for 12 h (see SI for details). The powder XRD pattern (Fig. 3b) displayed then relatively broad peaks unambiguously assigned to $BaNiS_2$[38] besides sharp peaks due to BaS and NiS present in 31 wt% and 28 wt%, respectively (see Table S3 and Fig. S9 for Rietveld refinement). Clearly, the decomposition of $BaS_2$ into BaS severely competed with the insertion of Ni. Indeed this parasitic reaction was bypassed by introducing a $BaS_3$:Ni = 1:2 mixture into the preheated furnace and subsequent annealing for 1h. The obtained XRD pattern indicates much better $BaNiS_2$:BaS molar ratio up to 89:11. Our follow-up experiment revealed the reaction of Ni with $BaS_3$ first lead to the decomposition into $BaS_2$ and $NiS_x$ species (see Fig. S10 for details). Then remaining Ni intercalates into the $BaS_2$ produced in situ, which is likely stabilized by the sulfur vapor pressure built in the sealed tube. The reaction between $BaS_3$ and Fe suffered also from the decomposition into BaS and FeS, and such decomposition was suppressed using a similar heating treatment (Fig 3d). Introduction of the mixture into a preheated furnace followed by annealing at 340 °C for 1h yielded BaFe$_2$S$_3$[39] at 91 wt% with FeS and Ba$_6$Fe$_8$S$_{15}$ as very minor phases (See Fig. S11 for Rietveld refinement).

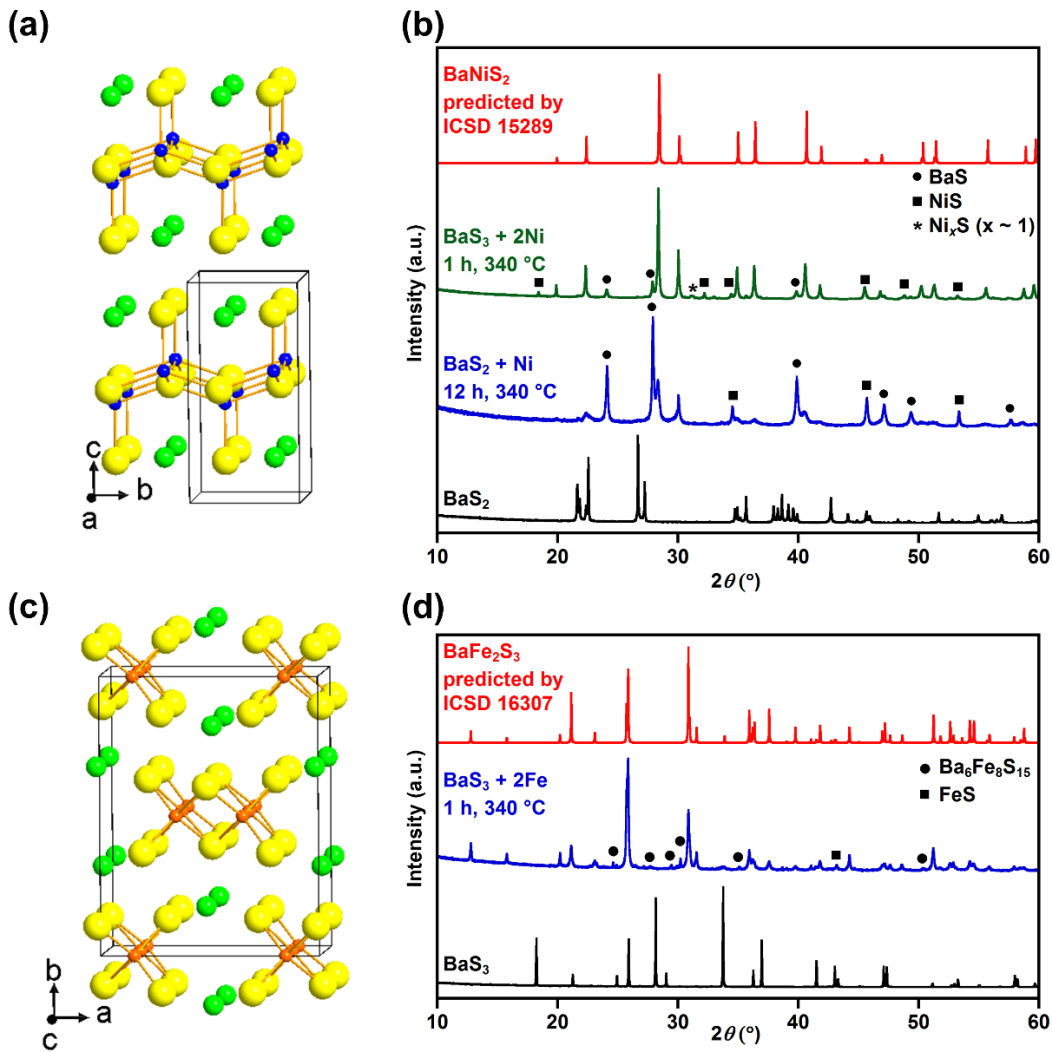

**Fig. 3.** (a) Structure of BaNiS$_2$; (b) theoretical XRD pattern of BaNiS$_2$ (red), experimental XRD pattern of the BaS$_3$ + 2Ni (green) and the BaS$_2$ + Ni (blue) mixture after thermal treatment at 340 °C for 1 and 12 h, respectively, as well as XRD pattern of BaS$_2$ (black). (c) Structure of BaFe$_2$S$_3$; (b) theoretical XRD pattern of BaFe$_2$S$_3$ (red), experimental XRD pattern of a BaS$_3$ + 2Fe mixture after the thermal treatment at 340 °C for 1 h (blue), and XRD pattern of BaS$_3$ (black).

These results demonstrate that the open-shell 3d transition metals (e.g. $Ni^{2+}$ and $Fe^{2+}$ in square-pyramidal and tetrahedral sites, respectively) can be inserted into $(S_n)^{2-}$ containing precursors at temperatures ($T$ = 340 °C) much lower than the ones ($T$ > 800 °C) needed for conventional ceramic routes for $BaNiS_2$ and $BaFe_2S_3$.[39-41]

In our previous report,[21] insertion of copper into $(S_2)^{2-}$ containing materials were shown to be topochemical. The same observation is made here as demonstrated below for $BaNiS_2$ and $BaFe_2S_3$ (Fig. 2f and 3d). The hypothetical topochemical reaction pathway for $BaCu_4S_3$ is also given in Fig. S12.

In $BaS_2$ (SG: $C2/c$), the $(S_2)^{2-}$ dimers occupy distorted octahedral sites of $Ba^{2+}$. In the case of $BaS_2 \rightarrow BaNiS_2$ transformation, small atomic rearrangement like the splitting of the dimers and concomitant separation of the basal planes of the $Ba_6$ octahedra are sufficient to reach the Ba-S framework of $BaNiS_2$ structure type. This creates a corrugated sulfur atomic layer that can then accommodate $Ni^{2+}$ cations (Fig. 4). Preservation of the structural building blocks and of the layered structure asserts the topochemical nature of the process.

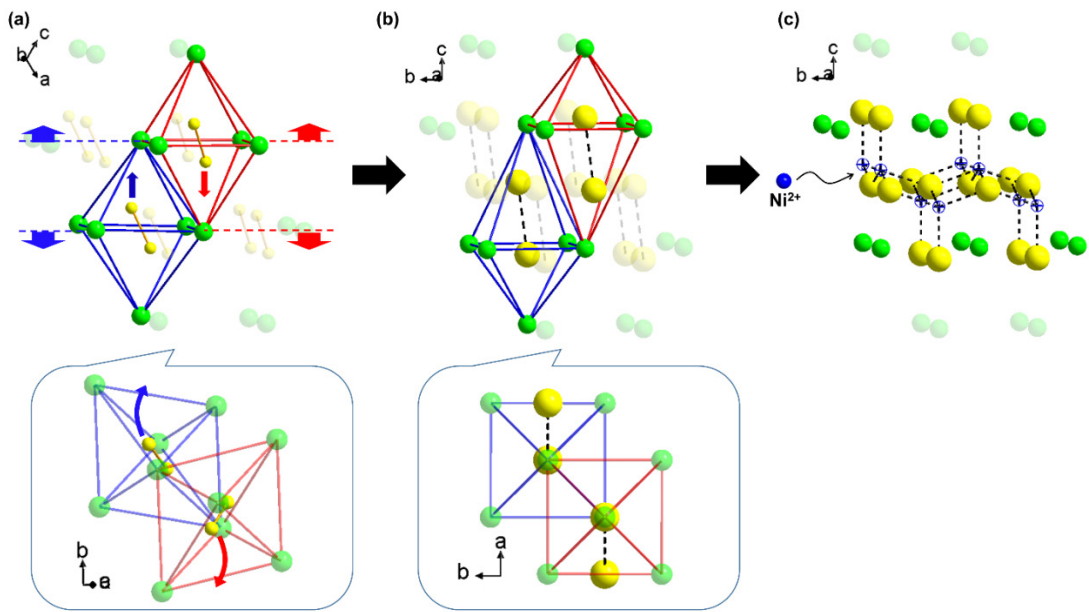

**Fig. 4**. Plausible reaction pathway from $BaS_2$ to $BaNiS_2$. (a) Structure of $BaS_2$. Two differently distorted octrahedra of $Ba_6S_2$ are colored in red and blue, respectively (Inset: projection of these two octahedra on their basal planes). (b) Structure of $BaNiS_2$, represented without the Ni atoms, highlighting the elongated $Ba_6$ octahedra and broken S-S pairs. (c) Structure of $BaNiS_2$ signifying the square-pyramidal sites for $Ni^{2+}$ cations and Ni-S bonds by blue hollow spheres and broken lines, respectively.

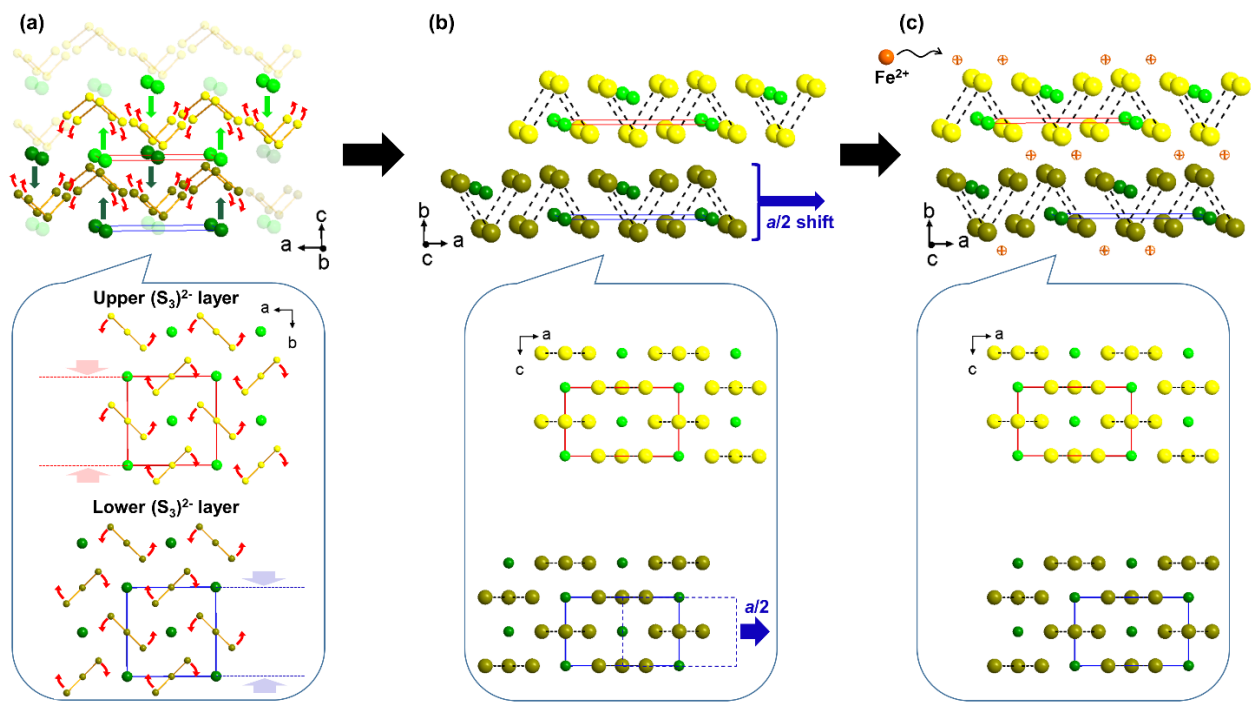

**Fig. 5**. Plausible reaction pathway from BaS$_3$ to BaFe$_2$S$_3$. A red or blue arrow describe movement of each structural unit. (a) Structure of BaS$_3$. Ba$^{2+}$ cations colored lime/deep green are merged into upper/lower (S$_3$)$^{2-}$ double layers colored yellow/brown, respectively. Concomitant S-S bond breaking leads (b) a hypothetical structure before the transformation to reach the structure of BaFe$_2$S$_3$ through $a$/2 shift of bottom $^2/_\infty$[BaS$_3$] layer. (c) Structure of BaFe$_2$S$_3$ displaying tetrahedral sites for Fe$^{2+}$ cations by orange hollow spheres.

In BaS$_3$ (SG: $P$-42$_1m$), a double layer made of two columns of (S$_3$)$^{2-}$ trimers is sandwiched within a base-centered sub-lattice of Ba$^{2+}$. In the case of BaS$_3$ → BaFe$_2$S$_3$ transformation, the conformational change of the broken trimers and concomitant segregation of Ba$^{2+}$ cations forms a [BaS$_3$]$^{2-}$ layer with the same framework as observed in the BaFe$_2$S$_3$ structure type (Fig. 5). The iron intercalation process occurs therefore between two successive [BaS$_3$]$^{2-}$ layers achieved through slight reorganization of the BaS$_3$ framework.

In both BaNiS$_2$ and BaFe$_2$S$_3$ cases the structural transformation from the precursor occurs without destructive reorganization of the precursor framework, which explains the low energy costs of the reactions. It supports an underlying topochemical nature of the processes as already observed in our previous study.

To sum up, we demonstrated the low-temperature reactivity of compounds containing (S$_n$)$^{2-}$ ($n$ = 2-3) oligomers, *i.e.* BaS$_2$ and BaS$_3$, towards transition metals to give BaCu$_2$S$_2$, BaCu$_4$S$_3$, BaNiS$_2$ and BaFe$_2$S$_3$ retaining their original structural motif. These reactions can be regarded as a thermo-assisted intercalation process involving the diffusion of the transition metal from the surface to the precursor bulk and the concomitant cleavage of S-S bonds thanks to electron transfer from the metal. The potential of this novel soft chemistry route for inorganic solids is tremendous according to the large number of precursors containing chalcogenide oligomers and might lead to the discovery of layered transition metal chalcogenides with remarkable properties.

## ACKNOWLEDGEMENTS

The authors thank P.-E. Petit and J.-Y. Mevellec in IMN for their help on X-ray diffraction measurements and Raman spectroscopy, respectively. S.S. is financially supported by JSPS Overseas Research Fellowships.

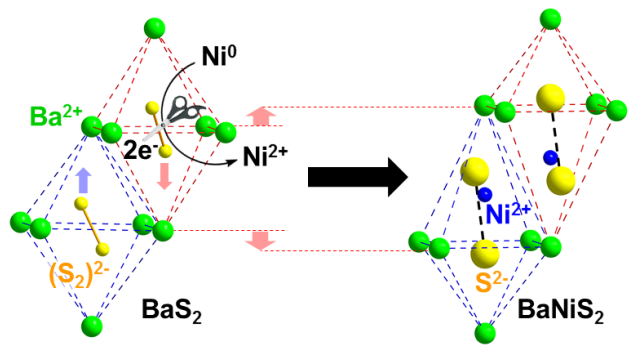

(For Table of Contents Only)